\documentclass[twocolumn,showpacs,preprintnumbers,amsmath,amssymb]{revtex4}
\usepackage{CJK}
\usepackage{graphicx}
\usepackage{dcolumn}
\usepackage{bm}
\usepackage{color}

\newcommand{\cO}{{\cal O}}
\newcommand {\eea}{\end{eqnarray}}

\newcommand{\vc}{\left({v_\infty\over c}\right)}
\newcommand{\cv}{\left({c\over v_\infty}\right)}
\newcommand{\post}{I have assumed, and will verify {\it a posteriori}, that }

\newcommand{\vtyp}{ 30{\rm{km}\over\rm{sec} }}

\newcommand {\gw}{GW150914 }

\begin{document}
\preprint{APS/123-QED}

\title{ Gravitational Radiation Assisted Capture}
\author{John Toner}
\affiliation{Department of Physics and Institute of Theoretical
Science, University of Oregon, Eugene, OR 97403}
\date{\today}
\begin{abstract}
It is shown that gravitational radiation can bind two  initially  unbound bodies; no third body is needed. Such captured  bodies will  almost always inspiral and merge due to further gravitational radiation on  cosmologically negligible time scales (e.g., @ 5 years for GW150914).  The capture cross-section $\sigma$ for such "capture and inspiraling" is far larger, for initial relative speed  of the two objects $v_\infty\ll c$, than that $\sigma_{d}$ for "direct capture":  
$\sigma\propto\left(c / v_\infty\right)^{18/7}$, while $\sigma_{d}\propto\left(c / v_\infty\right)^2$. Implications of these results for black hole binary mergers, and giant black holes at galactic centers, are discussed.
 \end{abstract} 
\maketitle

The direct detection of gravitational radiation\cite{GWD1, GWD2}, in addition to confirming one of the most important predictions of general relativity\cite{mtw}, raises the question of the origin of the black hole binaries that have been the source of both definite detections, and one possible detection, so far. Prior work\cite{3 body} has focussed on three-body mechanisms which bind two previously gravitationally unbound bodies. Such mechanisms have the disadvantage that their rate is proportional to the cube  of the density of stellar mass objects available to provide the third body, and is, therefore, very low when the density is low.

The purpose of this paper is to point out that gravitational radiation itself provides a very effective mechanism for {\it two body capture}, particularly of black holes\cite{Will}. I find that if two objects  of masses $m_1$ and $m_2$ with total mass $M\equiv m_1+m_2$ approach each other at asymptotic relative speed $v_\infty\ll c$, they will lose enough energy to be captured  into a highly elliptical orbit  if their impact parameter $b$ is less than a critical value $b_c$ given by:
\begin{eqnarray}
b_c&=&C_b\left(f(1-f)\right)^{1\over 7}\left({c \over v_\infty}\right)^{9/7}r_S(M)\nonumber\\&=&4.75\times10^5{\rm{km}}\left(f(1-f)\right)^{1\over 7}\left({M\over M_\odot}\right)\left({30{\rm{km}\over\rm{sec}}\over v_\infty}\right)^{9/7}\,
 \label{bc}
\end{eqnarray}
where $M_\odot$ is the mass of the Sun, $C_b\equiv \left({85\pi \over96}\right)^{1\over 7}\approx1.157367$, $f\equiv {m_1\over M}$, and  the Schwarzschild radius  $r_S(M)
={2GM/c^2}$. For GW150914, $M=65M_\odot$ (which implies $r_S(M)=192\rm{km}$) and  $f=29/65$,  assuming an initial relative velocity of $v_\infty=30{\rm{km}\over\rm{sec}}$ (a typical relative velocity for stars in the neighborhood of the Sun), 
equation (\ref{bc}) then gives
$b_c=2.53\times10^7\rm{km}=.17\rm{AU}$.

The cross-section for capture is given by
\begin{eqnarray}
\sigma=\pi b_c^2=C_\sigma[f(1-f)]^{2\over 7}\left({c \over v_\infty}\right)^{18/7}r_S^2(M),
\label{cross}
\end{eqnarray}
where  $C_\sigma=\left({7225\pi^9\over 9216}\right)^{1\over 7}\approx4.208$. Again for \gw with $v_\infty=30{\rm{km}\over\rm{sec}}$, this gives $\sigma=2\times10^{15}\rm{km}^2=.09\rm{AU}^2$.

Once captured in this way, the bodies will inspiral due to further gravitational radiation, until they  merge.
I find that the total inspiral time $\tau$ of the pair after  first periastron is 
\begin{eqnarray}
\tau&=&\pi\left({c^2 r_S(M)\over v_\infty^3}\right)\left({b\over b_c}\right)^{21/2}\zeta\left({3\over 2}, x\right)\nonumber\\&\approx&(1{\rm year})\left({30{\rm{km}\over\rm{sec}}\over v_\infty}\right)^{3}\left({M\over M_\odot}\right)\left({b\over b_c}\right)^{21/2}\zeta\left({3\over 2}, x\right), \nonumber\\
\label{inspiral}
\end{eqnarray}
where I've defined $x\equiv1-\left({b\over b_c}\right)^{7}$, and
\begin{eqnarray}
\zeta\left(y, x\right)\equiv \sum_{n=0}^{\infty} {1\over (x+n)^y}
\nonumber
\end{eqnarray}
is the Hurwitz zeta function\cite{hurwitz}.

Note that inspiral time $\tau$  is much less than the age of the universe for any reasonable mass $M$ and relative velocity $v_\infty$; for example, for GW150914, assuming  the impact parameter takes on its median value $b=b_c/\sqrt{2}$, and, as before, taking $v_\infty=30{\rm{km}\over\rm{sec}}$, I  obtain $\tau\approx4.80$ years. Therefore, virtually all pairs of masses  captured in this way will inspiral essentially instantaneously on a cosmological timescale. As a result, the limit on the rate of mergers caused by this
mechanism is the capture rate, not the subsequent inspiral.  Only the extremely rare occurrence of an impact parameter $b$ extremely close to $b_c$ can lead to cosmologically significant inspiral times.
For example, for \gw 
with all of the assumptions made earlier, achieving  $\tau\sim 10^9\, \rm{years}$ would require that $1-{b\over b_c}\le2.3\times 10^{-6}$; the probability of this is only $4.6\times 10^{-6}$. The distribution of inspiral times is very broad, however, as I will discuss further in the SM\cite{SM}.

One experimental signature of this mechanism of capture is that it would lead to inspiraling orbits of detectable eccentricity. For a given $v_\infty$, once the bodies have inspiraled to an elliptical orbit with periastron  $r_p\ll r_{p0}$, the distribution of eccentricities of the inspiraling pair  implied by this mechanism is:
\begin{eqnarray}
p(e;r_p)=
C_p\left({r_p\over r_S(M)}\right)[f(1-f)]^{-{2\over 7}}\left({v_\infty\over c}\right)^{4\over 7} e^{-{31\over19}}\,,
\label{edist}
\end{eqnarray}
 for $e>e_{min}$ and $p(e;r_p)=0$ for $e<e_{min}$, where
\begin{eqnarray}
e_{min}=C_{em}\left({r_p\over r_S(M)}\right)^{19\over12}[f(1-f)]^{-{19\over 42}}\left({v_\infty\over c}\right)^{19\over 21} \,\,,
\label{emin}
\end{eqnarray}
$C_p={6\over19}\left({425\over304}\right)^{{870\over2299}}\left({96\over85\pi}\right)^{{2\over7}}=.2676222338...$, and
\newline
\newline
$C_{em}=\left({425\over304}\right)^{{145\over242}}\left({96\over85\pi}\right)^{{19\over42}}2^{-{19\over12}}=
.25678305711...$. \newline

To predict the actual distribution of observations, for which the asymptotic approach velocity $v_\infty$ will of course be unknown, this must be averaged over the distribution of approach velocities. Doing this for a Maxwellian distribution of speeds with variance $v_{\sigma}^2$, I find\cite{SM}
\begin{eqnarray}
p(e)=
\bar{C}\left(1\over e_c\right)\left(e_c\over e\right)^{{31\over19}}\gamma\left({25\over14},{1\over 2}\left({e\over e_c}\right)^{42\over19}\right)\,,
\label{pevbarf}
\end{eqnarray}
where $\gamma(s, x)$ is the lower incomplete gamma function\cite{hurwitz}, I've defined the characteristic eccentricity scale 
\begin{eqnarray}
e_c=C_{em}\left({r_p\over r_S(M)}\right)^{19\over12}[f(1-f)]^{-{19\over 42}}\left({v_\sigma\over c}\right)^{19\over 21} \,\,,
\label{ec}
\end{eqnarray}
and the constant $\bar{C}\equiv2^{2\over7}\left({24\over19\sqrt{\pi}}\right)=.8687429...$.

The probability distribution (\ref{pevbarf}) has the limiting forms:
\begin{eqnarray}
p(e)\approx
\left\{\begin{array}{ll}
{C_<\over e_c}\left({e\over e_c}\right)^{44\over19},& e\ll
e_c\,,\\ {C_>\over e_c}\left({e_c\over e}\right)^{31\over19},& e\gg
e_c\,,
\end{array}\right.
\label{pelim} 
\end{eqnarray}
where I've defined $C_<\equiv{168\over475\sqrt{2\pi}}=.141099585...$ and $C_>\equiv\bar{C}\Gamma\left({25\over14}\right)=.805906...$, where $\Gamma$ is the {\it complete} Gamma function. The probability $p(e)$ has a single maximum of $p_{max}=.190385868/e_c$ at $e=1.90367666 e_c$.

The $e\gg e_c$ asymptotic scaling $p(e)\propto e^{-{31\over19}}$ holds for {\it all} distributions of the asymptotic speed 
$v_\infty$, with the replacements $v_\sigma\rightarrow v_c$ in  (\ref{ec}), and  $C_>\rightarrow{12\over19}$ in (\ref{pelim}), where 
$v_c\equiv\left(\left<v_\infty^{4/7}\right>\right)^{7/4}$,
with the brackets denoting an average over speeds. 

Since these results  ignore relativistic effects, for comparison with observational data, they should be used 
at a value of $r_p$ sufficiently large compared to $r_S$ (say, $r_p\sim10r_S(M)$) that relativistic effects are negligible, and then use that eccentricity as an initial condition for a numerical solution for the final stages of the inspiral. 

Note that the {\it typical} eccentricities $e_c$ (eqn. (\ref{ec})) will be very small if the rms velocity variance $v_\sigma\ll c$. For example,  taking $v_\sigma=30{\rm{km}\over\rm{sec}}$, $f=29/65$  (the value for \gw), and $r_p=10r_S(M)$ gives $e_c=4.45\times 10^{-3}$. Nonetheless, if, as is anticipated\cite{frey}, LIGO  eventually detects hundreds of black hole binary mergers, my result (\ref{pevbarf}) implies that some of these will have appreciable eccentricities: e.g.,  $7\%$  of  all mergers will have an eccentricity greater than $100e_c$, which is $\sim.445$ for the parameter values just assumed. This should be detectable.

This gravitational radiation assisted capture mechanism (hereafter "GRAC") may dominate 
the creation of both binary black hole mergers, and 
supermassive black holes (hereafter "GBH's") at galactic centers.

For both processes,  there are well-defined limits in which GRAC becomes infinitely more  effective than the other two competing mechanisms: direct capture (that is,  the two objects plunging directly into each other on their first passage), and three body capture. The cross-section for direct capture for $f\ll1$ 
is $\sigma_d=4\pi r_S^2(M)\cv^2$\cite{SM}. While calculating the direct capture cross-section for objects of comparable mass would require numerical solution of the full equations of general relativity, it is presumably of this order of magnitude. Therefore, the ratio of this direct capture cross-section to that of the gravitational wave assisted mechanism I consider here (given by equation (\ref{cross})) is $\sim\vc^{ 4\over7} f^{-{2\over7}}$; hence, direct capture is much less common, in the limit       $v_\infty\rightarrow 0$, than GRAC.  

Note, however, that because its cross-section does not vanish as $f\rightarrow 0$, direct capture  surpasses GRAC 
for  $f\lesssim\vc^2$. This is clearly the case for, e.g., the 
capture of subatomic dark matter particles by  giant black holes at  galactic centers\cite{GBH}. On the other hand, for the capture of {\it stars} by a giant black hole, GRAC is more effective even if the black hole is enormous;  for example, for  the giant black hole at the center of our own galaxy\cite{Milky GBH}, $M\sim4\times10^6M_\odot$, even stars of mass $\sim M_\odot/10$ (i.e., the mass of a typical star), can satisfy the condition $f\gtrsim
\left({v_\infty\over c}\right)^2$   for  relative asymptotic speeds $v_\infty\sim{30{\rm{km}\over\rm{sec}}}$. So although dark matter is far more common, the principle component of the diet of giant black holes may be stars.

Note also that GRAC is actually much {\it more} effective in the early stage of the growth of such a giant black hole, since $f$ is much smaller at that stage (when the GBH is much lighter). This could potentially explain how 
giant black holes grow\cite{GBH2} from intermediate mass black holes.

For the formation of BH binaries, as noted earlier, this mechanism is always favored over three body mechanisms as the number density 
$\rho\rightarrow 0$, since three body 
rates vanish like $\rho^3$, whereas the rate for two body processes like GRAC vanish like $\rho^2$.

Note, however, that the rate for three body mechanisms can scale like $\rho^2$ if an $\cO(1)$ fraction of the black holes are formed in bound pairs\cite{3 body}.
Furthermore, in high density regions, not only is the three body rate faster, but GRAC becomes less effective, since the highly elliptical orbits created by this capture are quite delicate, and easily gravitationally perturbed by  a third body. 

These are clearly quantitative questions which should be investigated to determine how important a role GRAC  plays in the creation of binary BH mergers.


I'll now derive the above results. Detailed calculations are given in the Supplemental Materials\cite{SM}; here 
I will  give simple rough arguments that recover the above results up to numerical factors of $\cO(1)$.

Consider two bodies approaching each other at non-relativistic speeds ($v_\infty\ll c$) with impact parameter $b$.
For Newtonian motion, conservation of energy and angular momentum imply\cite{SM} that the distance of closest approach $r_{p0}$ of the two bodies on their first passage is:
\begin{eqnarray}
r_{p0}={v_\infty^2b^2\over 2GM}=\vc^2{b^2\over r_S(M)} \,\,
 \label{rmb}
\end{eqnarray}
where  \post
 $b\gg r_{p0}$ for all captured orbits. This condition also implies that the relative  speed $v(r_{p0})$ of the pair at closest approach is nearly the escape velocity at that distance: 
 \begin{eqnarray}
 v(r_{p0})\approx\sqrt{{2GM\over r_{p0}}}
 \label{vp0}
 \end{eqnarray}

With the parameters of the orbit in hand, we can now calculate the energy emitted by gravitational
radiation on the first passage.
To do so, 
I begin with the general expression\cite{mtw} for the power $P$ emitted by a weak, slow-moving ($v\ll c$) gravitational wave source:
\begin{eqnarray}
P={G\over 5c^5}\dddot{Q}_{ij}\dddot{Q}_{ij}\,\,,
\label{P1}
\end{eqnarray}
where 
\begin{eqnarray}
Q_{ij}\equiv \sum_{\alpha} m_\alpha\left(r_i^\alpha r_j^\alpha-{1\over 3} \delta_{ij}|{\bf r}_\alpha|^2\right)\,\,
\label{Q1}
\end{eqnarray}
is the usual mass quadrupole tensor of a set of masses labeled by $\alpha$. Here there are only two masses $m_1$ and $m_2$, which, in  center of mass coordinates,  are located at ${\bf r}_1=-{m_2\over M}{\bf r}$ and ${\bf r}_2={m_1\over M}{\bf r}$ respectively, where ${\bf r}\equiv{\bf r}_2-{\bf r}_1$ is the relative displacement of the two masses.

I will verify {\it a posteriori} 
that the assumptions of slow motion (i.e.,  $v\ll c$) and weak gravitational fields are valid for the initial 
capture, and most of the inspiral process, for almost all pairs captured by GRAC.  This 
means 
the orbits are nearly Newtonian\cite{Kepler}, which makes it possible to do all calculations  analytically. 

Using the center of mass coordinates for the two masses in (\ref{Q1}), a typical component of  the mass quadrupole
tensor $\mathbf{Q}$ can be estimated entirely in terms of ${\bf r}$:
\begin{eqnarray}
Q_{ij}\sim{m_1m_2\over M}r^2=\mu r^2\,,
\label{Q2}
\end{eqnarray}
where $\mu\equiv {m_1m_2\over M}$ is the usual reduced mass.
Taking three time derivatives of this expression near periastron, where most of the gravitational radiation occurs,  essentially amounts to multiplying it by $\omega^3$, where $\omega\equiv{v(r_{p0})\over r_{p0}}$ is the angular velocity of the pair at periastron. Using this and (\ref{vp0})  gives, near periastron,
\begin{eqnarray}
\dddot{Q}_{ij}\sim{G^{3\over2}m_1m_2M^{1\over2}\over {r^{5\over2}_{p0}}}\,\,.
\label{Qddd}
\end{eqnarray}
Using this in the general expression (\ref{P1}) gives, for  the emitted power at periastron:
\begin{eqnarray}
P_p\sim{G^4m_1^2m_2^2M\over c^5 r_{p0}^5}  \,.
\label{P2}
\end{eqnarray}
This power is emitted for a time $\delta t$ of order $\delta t\sim{r_{p0}\over v(r_{p0})}$; hence the total energy emitted on the first passage is 
\begin{eqnarray}
\Delta E= P_p\delta t\times\cO(1)=C_E {G^{7\over 2}m_1^2m_2^2M^{1\over 2}\over c^5 r_{p0}^{7\over 2}}\,.
\label{delE3}
\end{eqnarray}
The detailed analysis given in the supplemental materials\cite{SM} recovers precisely this result, with a numerical prefactor of  $C_E\equiv 85\pi\sqrt{2}/24\approx 15.73521$. Using my earlier expression (\ref{rmb}) for the distance of closest approach  $r_{p0}(b)$  in this estimate of 
$\Delta E$ gives
\begin{eqnarray}
\Delta E=D_E{G^7m_1^2m_2^2M^4\over c^5 b^7v_\infty^7} \,,
\label{delE2}
\end{eqnarray}
where a precise calculation\cite{SM} gives the numerical prefactor $D_E=170\pi/3\approx178$.
When this energy loss is greater than the total original Newtonian energy of the system, which is just the center of mass kinetic energy at infinity, the two masses will become bound. The largest impact parameter $b_c$ that  satisfies this condition therefore obeys 
\begin{eqnarray}
\Delta E(b_c)={m_1m_2\over 2M}v_\infty^2\,.
\label{bc1}
\end{eqnarray}
Combining (\ref{delE2}) with (\ref{bc1}), using the fact that $r_S(M)=2GM/c^2$, and solving for $b_c$, gives equation (\ref{bc}).
Using the fact that the  capture cross section $\sigma=\pi b_c^2$ then immediately gives my principal result, equation (\ref{cross}).

I can now verify {\it a posteriori} my earlier assumption  of slow motion (i.e.,  $v\ll c$) and weak gravitational fields by noting that both of these assumptions are satisfied if $r\gg r_S(M)$ throughout the orbit, which is clearly true if the periastron distance on first passage $r_{p0}\gg r_S(M)$. This is readily verified using  (\ref{bc}) for the maximum impact parameter $b_c$ and the relation (\ref{rmb}) between $r_{p0}$ and $b$, which, taken together, imply 
\begin{eqnarray}
r_{p0}=\cv^{4/7}\left({85\pi f(1-f)\over 96}\right)^{2\over7} \left(b\over b_c\right)^2r_s(M)\,,
\label{rmin2}
\end{eqnarray}
from which it is clear that $r_{p0}\gg r_S(M)$ if $v_\infty\ll c$, unless $f\ll 1$ or $b\ll b_c$. The latter condition will rarely happen. Hence,  the motion will be nearly Newtonian if $v_\infty\ll c\sqrt{f}$. This condition will be satisfied by any objects of roughly equal mass for $v_\infty\ll c$, and  by stars approaching giant black holes for relative velocities $\lesssim \vtyp$.

The inspiral time can now be calculated by assuming 
that each subsequent return of the pair to periastron will occur at almost exactly the same periastron distance: 
$r_{pn}\approx r_{p0}$ until $n\gg 1$. Furthermore, the orbit during this "constant $r_{pn}$" phase of the inspiral is nearly parabolic near periastron. Finally, almost of of the inspiral time is spent in this "constant $r_{pn}$" phase. These 
statements will all be verified {\it a posteriori} in the SM.

Since the
periastron distances $r_{pn}\approx r_{p0}$, and the orbit remains nearly parabolic near periastron,  the energy loss on each  return will be nearly the same as that on the first passage. Hence, the energy after $n$ orbits is given by
\begin{eqnarray}
E_n=E_0-n\Delta E\,,
\label{En}
\end{eqnarray}
where $E_0$ is the energy {\it after} the first passage.
My result (\ref{delE2}) for $\Delta E$ can be rewritten as
\begin{eqnarray}
\Delta E={\mu v_\infty^2\over2}\left({b_c\over b}\right)^{7}\,.
\label{DelE}
\end{eqnarray}
Using the standard relation\cite{Newton} between semi-major axis $a$  and energy then gives
\begin{eqnarray}
a_n=-{Gm_1m_2\over2E_n}={Gm_1m_2\over2(n\Delta E-E_0)}={Gm_1m_2\over2(n+x)\Delta E}\,,
\label{an}
\end{eqnarray}
where I've defined
\begin{eqnarray}
x\equiv-{E_0\over\Delta E}=-{\left({\mu v_\infty^2\over2}-\Delta E\right)\over\Delta E}=\left(1-\left({b\over b_c}\right)^{7}\right)  \,.
\label{E0}
\end{eqnarray}
Using my expression  (\ref{DelE}) for the energy loss per orbit $\Delta E$, 
I obtain
\begin{eqnarray}
a_n={1\over2}\left({c\over v_\infty}\right)^2\left({1\over n+x}\right)\left({b\over b_c}\right)^{7}r_S(M)\,,
\label{an2}
\end{eqnarray}
where I've used $r_S(M)=2GM/c^2$ again. Using the standard relation $T_n=2\pi\sqrt{{a_n^3\over GM}}$ for the period $T_n$ of the $n$'th orbit, and summing this from $n=0$ to infinity gives the total inspiral time  equation (\ref{inspiral}).

The results that $r_{pn}\approx r_{p0}$ and (\ref{an}) imply that a very large number of orbits will have $r_{pn}\ll a_n$; i.e., their eccentricities will be very close to 1. Indeed, I show in the SM\cite{SM} that this will be the case until 
\begin{eqnarray}
n\sim n_c=\cv^{10/7}[f(1-f)]^{-2/7}\left({b\over b_c}\right)^5\,\,,
\label{epsilon}
\end{eqnarray}
which is $\gg1$ for $v_\infty\ll c$ unless 
$b\ll b_c$, which is highly unlikely, or $f\ll\vc^5$, i.e., 
widely disparate masses. This  large value of $n_c$ justifies extending the sum to $n=\infty$ in the calculation of the total inspiral time (\ref{inspiral}).

To derive of the final eccentricity distribution law (\ref{edist}),
I begin with the equations for the evolution of the eccentricity $e$ and semi-major axis $a$ of an inspiraling, nearly Newtonian orbit derived by Peters\cite{Peters}:
\begin{eqnarray}
{da\over dt}=-{64K(1+{73\over24}e^2+{37\over96}e^4)\over 5a^3(1-e^2)^{^{7\over 2}}}\equiv-{64Kf(e)\over 5a^3(1-e^2)^{^{7\over 2}}}\,\,,
\label{adot}
\end{eqnarray}
\begin{eqnarray}
{de\over dt}=-{K(e(304+121e^2))\over 15a^4(1-e^2)^{^{5\over 2}}}\equiv-{Kg(e)\over 15a^4(1-e^2)^{^{5\over 2}}}\,\,.
\label{edot}
\end{eqnarray}
where I've defined $K\equiv G^3m_1m_2M/ c^5$.

These equations were derived by Peters\cite{Peters} in the approximation that the parameters $a$ and $e$ undergo only small percentage changes on each orbit. This is clearly {\it not} the case for $a$ for the first few orbits after capture, as inspection of my expression (\ref{an}) for the semi-major axis $a_n$
of the $n$'th orbit makes clear. 

However, 
as noted earlier, the distance of closest approach of the $n$'th orbit $r_{pn}$ does {\it not} vary appreciably from orbit to orbit (until the very latest stages of the inspiral, which contribute negligibly to the total inspiral time). Furthermore, as detailed in the SM\cite{SM}, by combining (\ref{adot}), (\ref{edot}), and the elementary relation $r_p=a(1-e)$, I obtain a differential equation describing the evolution of $r_p$ as a function of eccentricity $e$:
\begin{eqnarray}
{d\ln r_p\over de}={d\ln a\over de}-{1\over 1-e} ={y(e)\over e}\,\,,
\label{dlnrde}
\end{eqnarray}
where $y(e)$ is a rational function of e, given explicitly in the SM\cite{SM},  that is finite and $\cO(1)$ for all $e$ in the range $0\le e\le 1$.
Hence, I can use this differential equation out to $e=1$. Doing so, 
I find \cite{SM} that the solution of  (\ref{dlnrde}) with the initial conditions $e=1$ and $r_p=r_{p0}$ implies that by the time $e\ll1$,  
\begin{eqnarray}
e(r_p)=C_e\left({r_p\over r_{p0}}\right)^{19\over12}\,\,.
\label{rpe3}
\end{eqnarray}
where $C_e=\left({425\over304}\right)^{{145\over242}}/2^{{19\over12}}=.40790...$.
Using the relation (\ref{rmb}) between the minimum first passage distance $r_{p0}$ and the 
impact parameter $b$, I can rewrite this as a relation between the eccentricity and the impact parameter:
\begin{eqnarray}
e(b)=C_e\left({r_Sr_p\over b^2}\cv^2\right)^{19\over12}\,\,.
\label{eb}
\end{eqnarray}
Solving for $b(e)$ gives
\begin{eqnarray}
b(e)=C_e^{6\over19}\sqrt{r_Sr_p}\cv e^{-{6\over19}}\,\,.
\label{be}
\end{eqnarray}
The probability distribution for the final eccentricity can now be obtained from that for the impact parameter $b$ via simple statistics, which imply:
\begin{eqnarray}
p(e;v_\infty)=p(b;v_\infty)\left|{db\over de}\right|\,\,.
\label{pe1}
\end{eqnarray}
Since the impact parameters of captured pairs should be uniformly distributed over a circle of radius $b_c$, I have
\begin{eqnarray}
p(b;v_\infty)={2b\over b_c^2(v_\infty)}\,\,.
\label{pb}
\end{eqnarray}
Using this and (\ref{be}) in (\ref{pe1}) gives the probability distribution (\ref{edist}) for the final eccentricity. In the SM\cite{SM}, I show that averaging this over a Maxwellian speed distribution gives (\ref{pevbarf}). I also show in the SM that the $e\gg e_c$ limit of the velocity averaged distribution of final eccentricities (i.e., the second line of equation (\ref{pelim})) is universal for all speed distributions, in the sense described earlier.

I thank J. Brau and R. Frey for useful discussions, R. Zimmerman for encouragement, and the Marsh Cottage Institute for Astrophysics, Inverness, CA for its hospitality while a portion of this work was underway.

\newpage

\section{Supplemental Materials}
 
\subsection{Distance of closest first approach $r_{p0}$}

I will treat the motion as Newtonian, which can be justified {\it a posteriori} by showing that the distance of closest approach  on first passage $r_{p0}$ obeys $r_{p0}\gg r_S(M)$. I will also
assume that 
$r_{p0}\ll b$, the impact parameter. I'll later verify {\it a posteriori} that this holds for all pairs that are captured by GRAC.

Given this, it is clear from conservation of angular momentum, which implies 
\begin{eqnarray}v(r_{p0})r_{p0}=v_\infty b \,,
\label{Lcons}
\end{eqnarray}
that the speed $v(r_{p0})\gg v_\infty$. This in turn implies that most of the kinetic energy of the  pair at periastron is obtained from their potential energy. Of course, an orbit on which the  kinetic energy is equal to the potential energy is parabolic. Hence, the orbit near periastron, which is where most of the gravitational radiation will take place (as we'll see below), is nearly parabolic. The velocity at periastron is therefore very close to the escape velocity at that radius; hence
\begin{eqnarray}
v(r_{p0})=\sqrt{{2GM\over r_{p0}}} \,.
\label{vrpSM}
\end{eqnarray}
Using this in (\ref{Lcons}) and solving for $r_{p0}$ gives 
\begin{eqnarray}
r_{p0}={v_\infty^2b^2\over 2GM}=\vc^2{b^2\over r_S(M)} \,\,,
 \label{rmbSM1}
\end{eqnarray}
which is just equation (\ref{rmb}).

I'll now verify my two {\it a posteriori} assumptions above. First, to see that $r_{p0}\ll b$, 
I take the ratio ${r_{p0}\over b}$ using (\ref{rmbSM1}), which gives ${r_{p0}\over b}=\vc^2{b\over r_S(M)}$. Since this ratio grows with increasing impact parameter $b$, its {\it largest} possible value for a captured pair occurs when $b=b_c$.  Hence, ${r_{p0}\over b}\le\vc^2{b_c\over r_S(M)}=C_bf(1-f)\vc^{5\over7}$, where in the second equality I have used (\ref{bc}) of the main text for $b_c$. Note that this ratio is clearly much less than $1$ if $v_\infty\ll c$. (Recall that $C_b=\left({85\pi \over96}\right)^{1\over 7}\approx1.157367$ and $f(1-f)\le1/4$).

To see that $r_{p0}\gg r_S(M)$, I consider the ratio ${r_{p0}\over r_S(M)}=\vc^2\left({b\over r_S(M)}\right)^2=\vc^2\left({b_c\over r_S(M)}\right)^2\left({b\over b_c}\right)^2$. 
Again using my result (\ref{bc}) for the maximum impact parameter for capture $b_c$, I find
${r_{p0}\over r_S(M)}=\cv^{4\over7}C_b^2\left(f(1-f)\right)^{2\over7}\left({b\over b_c}\right)^2$, which will always be much greater than $1$ for $v_\infty\ll c$ unless $b\ll b_c$, which will very rarely happen, or $f\ll\vc^2$, which can only happen if the two bodies are of extremely disparate masses; even a brown dwarf with $M\sim {M_\odot\over10}$ encountering the GBH at the center of our galaxy ($M_{GBH}\sim {4\times10^6}M_\odot$) will violate this condition.

\subsection{Constancy of $r_{pn}$ for most of the inspiral time}

I begin by considering the first passage. In the main text, I have already estimated the energy loss (\ref{delE2}) on this passage. The angular momentum loss rate $\dot{\bf{L}}$ can also be expressed in terms of the mass quadrupole tensor of the pair\cite{mtw}:
\begin{eqnarray}
\dot{L}_i=-{G\over 5c^5}\epsilon_{ijk}\ddot{Q}_{jm}\dddot{Q}_{km}\,\,.
\label{L1}
\end{eqnarray}
As I did in the main text for the energy, I can estimate the total change $\delta L$ in the magnitude of the angular momentum on the first passage by replacing each of the five time derivatives in (\ref{L1}) with the angular velocity $\omega\sim{v_p\over r_p}$, estimating $Q$ itself by $\mu r_{p0}^2$, where $\mu$ is the reduced mass, and multiplying the resultant rate by the rough estimate $\delta t\sim {r_{p0}\over v(r_{p0})}$ of the time spent near periastron on the first passage. Doing this gives
\begin{eqnarray}
\delta L\sim-{G\over c^5}\left({v(r_{p0})\over r_{p0}}\right)^5 \mu^2 r_{p0}^4\left({r_{p0}\over v(r_{p0})}\right)=-{G\mu^2v^4(r_{p0})\over c^5}\,\,.\nonumber\\
\label{del L1}
\end{eqnarray}
Using my earlier result (\ref{vrpSM}) for $v(r_{p0})$ in this expression gives
\begin{eqnarray}
\delta L\sim-{G^3\mu^2M^2\over c^5r_{p0}^2}\,\,.
\label{del L2}
\end{eqnarray}
Now using eqn. (\ref{rmbSM1}) to relate $r_{p0}$ to the impact parameter $b$, and using 
$\mu=f(1-f)M$ and ${2GM\over c^2}=r_S(M)$, I can rewrite this as
\begin{eqnarray}
\delta L\sim-Mr_S(M)c(f(1-f))^2\cv^4\left({r_S(M)\over b}\right)^4\,\,.
\label{del L3}
\end{eqnarray}
This is a small fraction of of the initial center of mass angular momentum $L_0=\mu v_\infty b=f(1-f)M v_\infty b$ of the pair, as can be seen by taking the ratio:
\begin{widetext}
\begin{eqnarray}
{|\delta L|\over L_0}\sim f(1-f)\cv^5\left({r_S(M)\over b}\right)^5=f(1-f)\left({b_c\over b}\right)^5\cv^5\left({r_S(M)\over b_c}\right)^5\,\,.
\label{del L4}
\end{eqnarray}
\end{widetext}
Using my expression (\ref{bc}) for the critical impact parameter $b_c$ in this expression gives
\begin{eqnarray}
{|\delta L|\over L_0}\sim( f(1-f))^{2\over7}\vc^{10\over7}\left({b_c\over b}\right)^5\,\,.
\label{del L5}
\end{eqnarray}
Since $f(1-f)<1$, this ratio will always be much less than $1$ if $v_\infty\ll c$, unless the impact parameter $b\lesssim b_c\vc^{2\over7}$, which will rarely happen (indeed, the probability of it happening  is $\left(b\over b_c\right)^2\sim \vc^{4\over7}=5\times 10^{-3}$ for $v_\infty=30{\rm{km}\over\rm{sec}}$).\newline

So the magnitude of the angular momentum $L$ after the first passage is almost the same as that before the first passage. 

I can determine the distance of closest approach $r_{p1}$ on the {\it second} passage using the fact that energy and angular momentum will be conserved until the next close passage. This implies 
that
\begin{eqnarray}
E_0={\mu v^2(r_{p1})\over2}-{GM\mu\over r_{p1}}\,\,
\label{econ}
\end{eqnarray}
and
\begin{eqnarray}
\mu v(r_{p1}) r_{p1}=L_{p1}\approx L_0=\mu v_\infty b\,\,,
\label{Lcon}
\end{eqnarray}
where in the second, approximate equality, I have used the result just derived that the angular momentum hardly changes between the first and the second passage.

If I assume, as I'll verify {\it a posteriori}, that $E_0$ is negligible compared to ${GM\mu\over r_{p1}}$, and solve (\ref{econ}) and (\ref{Lcon}) for $r_{p1}$, I get
\begin{eqnarray}
r_{p1}={v_\infty^2b^2\over 2GM}=\vc^2{b^2\over r_S(M)}=r_{p0} \,\,.
 \label{rmb1SM2}
\end{eqnarray}
Thus, the periastron distance
$r_{p1}$ of the second passage is, as I claimed in the main text, almost exactly equal to that of the first passage, provided I can verify my {\it a posteriori} assumption about the negligibility of the energy $E_0$.

To verify my  assumption that $E_0\ll{GM\mu\over r_{p1}}$, I take the ratio 
\begin{eqnarray}
\Gamma\equiv{|E_0|\over\left[{GM\mu\over r_{p1}}\right]}\,.
\end{eqnarray}
The magnitude $|E_0|$ of $E_0$ is bounded above by $\Delta E$, eqn. (\ref{delE2}), the energy loss on the first passage, since the initial energy $E_0$ of the objects before the first passage (i.e., as they approach from infinity) is positive, and $E_0=E_0-\Delta E$ is negative (since the captured orbit is bound). Hence
\begin{eqnarray}
\Gamma\le{\Delta E\over\left[{GM\mu\over r_{p0}}\right]} \,\,,
 \label{E0neg2}
\end{eqnarray}
Using
(\ref{rmbSM1}) and (\ref{delE2}) to relate $r_{p0}$ and $\Delta E$ to the impact parameter $b$, this expression can trivially be rewritten in terms of the maximum impact parameter for capture $b_c$ as
\newpage 
\begin{widetext}
\begin{eqnarray}
\Gamma\le{D_E\over 2} f(1-f)\left({GM\over cbv_\infty}\right)^5\le {D_E\over 2^{6}} \left({r_S(M)c\over bv_\infty}\right)^5f(1-f)<3\left({r_S(M)c\over b_cv_\infty}\right)^5\left({b_c\over b}\right)^5\left(f(1-f)\right)^{2\over7} \,,
\label{E0neg3}
\end{eqnarray}
\end{widetext}
where I've used the facts that $f(1-f)\le{1\over4}$ and $D_E<2^{8}$.

Using my expression (\ref{bc}) of the main text for the maximum impact parameter, this can be rewritten as
\begin{eqnarray}
\Gamma\le [f(1-f)]^{{2\over7}}\vc^{10\over7}\left({b_c\over b}\right)^5\,\,,
 \label{E0neg4}
\end{eqnarray}
which is clearly much less than $1$ for $v_\infty\ll c$ unless  $b\ll b_c$, which is very unlikely. 

Thus, in almost all cases, $E_0$ is negligible in determining $r_p$, which was my a posteriori assumption.

I can now repeat this argument for the third passage. The energy and angular momentum losses on the second passage will be almost the same as those on the first passage, since $r_p$ is virtually the same, and the orbit remains nearly parabolic near periastron (as illustrated by the negligibility of the energy.  Hence, the third periastron, and so on, will also be at the same distance, and result in the same losses of energy and angular momentum. 

The energy and angular momentum losses, and $r_{pn}$ itself, will only start to change when  the eccentricity of the orbit starts to be appreciably different from $1$. I can estimate how many orbits $n_c$ must occur before this happens by noting that $1-e_n= {a_n\over r_{pn}}$, where $e_n$ is the eccentricity of the $n$'th orbit. Using 
my expression (\ref{an2}) of the main text for $a_n$, taking $r_{pn}\approx r_{p0}$ with $r_{p0}$ given by (\ref{rmbSM1}), and considering $n\gg 1$, I obtain
\begin{eqnarray}
1-e_n={r_{pn}\over a_n }\approx{r_{p0}\over a_n }\approx2n\vc^{4}\left({b_c\over b}\right)^7\left({b\over r_S(M)}\right)^{2}\,\,.\nonumber\\
\label{1-eSM}
\end{eqnarray}
Using my expression (\ref{bc}) of the main text for $b_c$, and reorganizing, I can rewrite this as
\begin{eqnarray}
1-e_n\sim n\vc^{10\over7}[f(1-f)]^{2/7}\left({b_c\over b}\right)^5\,\,,\nonumber\\
\label{ncSM}
\end{eqnarray}
which is clearly $\ll1$ until $n\gtrsim n_c$, where
\begin{eqnarray}
 n_c=\cv^{10/7}[f(1-f)]^{-2/7}\left({b\over b_c}\right)^5\,\,,
\label{ncSM}
\end{eqnarray}
\newline
which is very large unless 
either $f\ll \vc^5$, which is not even satisfied for relative velocities at infinity $v_\infty=30{\rm{km}\over\rm{sec}}$ 
\newline 
for very small stars falling into the GBH at the center of our galaxy, or ${b\over b_c}\ll \vc^{2\over7}$, which will very rarely occur.

Since the sum over orbit number $n$ that enters the calculation of the inspiral time in the main text (i.e., the sum in the Hurwitz zeta function) converges as $n\rightarrow\infty$, and the value $n_c$ of $n$ at which my approximations break down is so large, it is quite accurate to extend this sum all the way out to $n=\infty$, as I have done in writing (\ref{inspiral}).

\subsection{Precise calculation of the the energy loss $\Delta E$ per passage}

Using the center of mass coordinates for the two masses in (\ref{Q1}), I can express the mass quadrupole
tensor $\mathbf{Q}$ entirely in terms of ${\bf r}$:
\begin{eqnarray}
Q_{ij}={m_1m_2\over M}\left(r_i r_j-{1\over 3} \delta_{ij}|{\bf r}|^2\right)\,\,.
\label{Q2SM}
\end{eqnarray}
Taking {\it two} time derivatives of this expression gives 
\begin{eqnarray}
\ddot{Q}_{ij}={m_1m_2\over M}\left[a_i r_j+a_jr_i+2v_iv_j-{2\over 3} \delta_{ij}\left({\bf a}\cdot{\bf r}+v^2\right)\right]\,\,,\nonumber\\
\label{Qdd}
\end{eqnarray}
where ${\bf v}\equiv \dot{\bf r}$ and ${\bf a}\equiv\ddot{\bf r}$ are the relative velocity and acceleration of the two masses. Using the equation of motion ${\bf a}=-{GM\over r^3}{\bf r}$ for ${\bf r}$, I can rewrite this expression as
\begin{widetext}\begin{eqnarray}
\ddot{Q}_{ij}=-{2Gm_1m_2\over r^3}\left(r_i r_j-{1\over 3} \delta_{ij}|{\bf r}|^2\right)+{2m_1m_2\over M}\left(v_i v_j-{1\over 3} \delta_{ij}|{\bf v}|^2\right)\,\,.
\label{Qdd2}
\end{eqnarray}
\end{widetext}
Now taking one further time derivative to obtain $\dddot{\bf Q}$, I obtain, after using the equation of motion for ${\bf r}$ again, 
\begin{widetext}
\begin{eqnarray}
\dddot{Q}_{ij}={2Gm_1m_2\over r^3}\left[-2\left(v_i r_j+v_jr_i\right)+{\bf v}\cdot{\bf r}\left({3r_ir_j\over r^2}+{\delta_{ij}\over 3} \right)\right]\,\,.
\label{QdddSM}
\end{eqnarray}
\end{widetext}

Inserting this into the general expression (\ref{P1}) for the emitted power P gives, after a little (!) algebra, 
\begin{eqnarray}
P={8G^3m_1^2m_2^2\over 5c^5 r^6}\left[4v^2r^2-{11\left({\bf v}\cdot{\bf r}\right)^2\over 3}\right]
\label{P2SM}
\end{eqnarray}

To proceed further, I use the fact that, for a Newtonian orbit,  the velocity vector ${\bf v}$ at any point on the orbit can be written as
\begin{eqnarray}
{\bf v}={\bf B}+{GM\over h}\hat{{\boldsymbol \theta}}
\label{v1}
\end{eqnarray}
where
the
"binormal" vector\cite{binormal} ${\bf B}$
is a constant of Newtonian motion which lies in the plane of the orbit perpendicular to the line from the origin to periastron, $h$ is the angular momentum about the center of mass divided by the reduced mass 
$\mu \equiv{m_1m_2\over M}=f(1-f)M$, and $\hat{{\boldsymbol \theta}}$ is the unit vector orthogonal to ${\bf r}$ in the plane of the orbit. 

For a parabolic orbit, 
\begin{eqnarray}
v=\sqrt{{2GM\over r}}
\label{v2}
\end{eqnarray}
everywhere. Applying this at periastron ($r=r_p$), where ${\bf v}$ is perpendicular to ${\bf r}$, so that $h=\left|{\bf v}\times {\bf r}\right|=vr$, I obtain
\begin{eqnarray}
h=r_p\sqrt{{2GM\over r_p}}=\sqrt{2GMr_p}\,.
\label{h}
\end{eqnarray}
Using this and (\ref{v2}) in (\ref{v1}), again applied at periastron, gives
\begin{eqnarray}
\sqrt{{2GM\over r_p}}\hat{{\boldsymbol y}}={\bf B}+\sqrt{{GM\over 2 r_p}}\hat{{\boldsymbol y}}\,,
\label{B1}
\end{eqnarray}
where I've defined my $x$ and $y$ axes to lie in the plane of the orbit along and perpendicular to the line to the periastron, respectively. Equation (\ref{B1}) can of course easily be solved for ${\bf B}$:
\begin{eqnarray}
{\bf B}=\sqrt{{GM\over 2 r_p}}\hat{{\boldsymbol y}}\,.
\label{B2}
\end{eqnarray}

Using polar coordinates in the orbital plane with $\theta=0$ along the $x$-axis, and using (\ref{v1}) and (\ref{B2}) for ${\bf v}$ and ${\bf B}$
respectively, I can write 
\begin{eqnarray}
{\bf v}\cdot{\bf r}={\bf B}\cdot{\bf r}=\sqrt{{GM\over 2 r_p}}r\sin\theta\,.
\label{B3}
\end{eqnarray}

Using this and (\ref{v2}) for the speed $v$ in my expression (\ref{P2SM}) for the power, I get
\begin{eqnarray}
P={8G^4m_1^2m_2^2M\over 5c^5 r^5}\left[8-{11 r\over 6r_p}\sin^2\theta\right]\,.
\label{P3}
\end{eqnarray}
I can now get the total power lost on the first passage by integrating this over all time:
\begin{eqnarray}
\Delta E&=&\int_{-\infty}^{\infty} P(t) dt \nonumber\\&=&{8G^4m_1^2m_2^2M\over 5c^5 }\int_{-\infty}^{\infty}{dt\over r^5}\left[8-{11 r\over 6r_p}\sin^2\theta\right]\,.
\label{delE1SM}
\end{eqnarray}
To evaluate the integral, I change variables of integration from time $t$ to angle $\theta$ using $d\theta/dt=h/r^2$, which follows from conservation of angular momentum. Using this in (\ref{P3}), and using the fact that, for a parabolic orbit, 
\begin{eqnarray}
r(\theta)={2r_p\over 1+cos\theta}\,,
\label{r(theta)}
\end{eqnarray}
together with my earlier expression (\ref{h}) for the angular momentum per unit reduced mass 
$h$, I obtain
\begin{widetext}
\begin{eqnarray}
\Delta E
={G^{7\over 2}m_1^2m_2^2M^{1\over 2}\over 5\sqrt{2}c^5 r_p^{7\over 2}}\int_{-\pi}^{\pi}d\theta\left[8(1+\cos\theta)^3-{11\over 3}(1+\cos\theta)^2\sin^2\theta\right]\,.
\label{delE2SM}
\end{eqnarray}
\end{widetext}
The angular integral in this expression is elementary. Evaluating it, I get
\begin{eqnarray}
\Delta E=C_E{G^{7\over 2}m_1^2m_2^2M^{1\over 2}\over c^5 r_p^{7\over 2}}
\label{delE3SM}
\end{eqnarray}
where $C_E\equiv 85\pi\sqrt{2}/24\approx 15.73521$. Using (\ref{rmbSM1}) to rewrite this in terms of the impact parameter $b$ gives
\begin{eqnarray}
\Delta E=D_E{G^7m_1^2m_2^2M^4\over c^5 b^7v_\infty^7} \,,
\label{delE4SM}
\end{eqnarray}
where $D_E=170\pi/3\approx178$.
When this energy loss is greater than the total original energy of the system, which is just the kinetic energy at infinity, the two masses will become bound. The largest impact parameter $b_c$ that will satisfy this condition is therefore given by solving
\begin{eqnarray}
\Delta E(b_c)={m_1m_2\over 2M}v_\infty^2\,.
\label{bc1SM}
\end{eqnarray}
Combining (\ref{delE4SM}) with (\ref{bc1SM}), using the fact that $r_S(M)=2GM/c^2$, and solving for $b_c$ gives equation (\ref{bc}).
Using the fact that the  capture cross section $\sigma=\pi b_c^2$ then immediately gives my principal result, equation (\ref{cross}).

\subsection{Derivation of the final eccentricity distribution law}

Finally, I turn to the derivation of the final eccentricity distribution law (\ref{edist}).
This begins with the equations for the evolution of the eccentricity and semi-major axis $a$ of an inspiraling, nearly Newtonian orbit derived by Peters\cite{Peters}:

\begin{eqnarray}
{da\over dt}=-{64Kf(e)\over 5a^3(1-e^2)^{^{7\over 2}}}\,\,,
\label{adotSM}
\end{eqnarray}
and the eccentricity $e$:
\begin{eqnarray}
{de\over dt}=-{Kg(e)\over 15a^4(1-e^2)^{^{5\over 2}}}\,\,,
\label{edotSM}
\end{eqnarray}
where I've defined $K\equiv G^3m_1m_2M/ c^5$,
\begin{eqnarray}
f(e)\equiv 1+{73\over24}e^2+{37\over96}e^4\,\,,
\label{fdefSM}
\end{eqnarray}
and
\begin{eqnarray}
g(e)\equiv e(304+121e^2)\,\,.
\label{gdefSM}
\end{eqnarray}

These equations were derived by Peters\cite{Peters} using a sort of adiabatic approximation, in which it is assumed that the parameters $a$, $e$, and $\epsilon\equiv1-e$ undergo only small percentage changes on each orbit. This is clearly {\it not} the case for $a$ for the first few orbits after capture, as inspection of my expression 

Taking the ratio of (\ref{adotSM}) and (\ref{edotSM}) gives a differential equation for the evolution of the semimajor axis $a$ with the eccentricity $e$:
\begin{eqnarray}
{da\over de}={\dot{a}\over\dot{e}} ={192af(e)\over (1-e^2)g(e)}\,\,,
\label{dadeSM}
\end{eqnarray}
a result also first obtained by Peters\cite{Peters}.
Since $a$ changes substantially between one orbit and the next for the first few orbits, I cannot actually use this differential equation for those orbits. It is therefore more useful
to rewrite this expression in terms of the periastron distance 
\begin{eqnarray}
r_p=a(1-e)\,,
\label{rpaSM}
\end{eqnarray}
which does {\it not} change substantially between orbits, even initially.

I can rewrite (\ref{dadeSM}) in terms of the periastron distance by first recasting it as an equation for $\ln a$:
\begin{eqnarray}
{d\ln a\over de}={192f(e)\over e(1-e^2)g(e)}\,\,,
\label{dlnadeSM}
\end{eqnarray}
and then using (\ref{rpaSM}) to write
\begin{eqnarray}
{d\ln r_p\over de}={d\ln a\over de}-{1\over 1-e} ={y(e)\over e}\,\,,
\label{dlnrdeSM}
\end{eqnarray}
where I've defined
\begin{eqnarray}
y(e)\equiv {192-112e+168e^2+47e^3\over(1+e)(304+121e^2)}\,\,.
\label{ydefSM}
\end{eqnarray}

The right hand side of equation (\ref{dlnrdeSM}) can be rewritten
\begin{eqnarray}
{y(e)\over e}={y(0)\over e}+{y(e)-y(0)\over e}= {12\over 19e}-z(e)\,\,.
\label{ysmalleSM}
\end{eqnarray}
where I've used the fact that   $y(0)={12\over 19}$, and 
\begin{widetext}
\begin{eqnarray}
z(e)\equiv {y(0)-y(e)\over e}={5776-1740e+559e^2\over19(1+e)(304+121e^2)}={1\over1+e}-{1740e\over19(121e^2+304)}\,\,,
\label{zdefSM}
\end{eqnarray}
\end{widetext}
is, by construction, finite as $e\rightarrow 0$. It is easy to check that $z(e)$ is in fact finite and $\cO(1)$ throughout the range $0\le e\le 1$, including the endpoints $e=0$ and $e=1$. I will make use of this fact in a moment.

Using (\ref{ysmalleSM}) in (\ref{dlnrdeSM}), and integrating
from the initial orbit to some  later orbit gives
\begin{eqnarray}
\ln \left({r_p\over r_{p0}}\right)={12\over19}\ln \left({e\over e_1}\right)-\int_{e_1}^e z(e')de'\,\,.
\label{rpe1SM}
\end{eqnarray}
For most captured orbits, the initial eccentricity $e_1$ is very close to $1$. Furthermore, the initial periastron distance $r_{p0}$ is much greater than the Schwarzschild radius $r_S$. Hence, if I wish to know the eccentricity when the pair has inspiraled  enough to emit detectable gravitational radiation, which only occurs when $r_p\sim r_S$, I need only consider $r_p\ll r_{p0}$. It can be seen from (\ref{rpe1SM}) that this implies that $e\ll1$. Using $e_1\approx1$ and $e\ll1$ in (\ref{rpe1SM}), and reversing the order of limits in the integral on the right  hand side so that the smaller value of $e$ is the lower limit, I see that, to an excellent approximation for most captured orbits,
\begin{eqnarray}
\ln \left({r_p\over r_{p0}}\right)={12\over19}\ln e(r_p)+\int_{0}^1 z(e')de'\,\,.
\label{rpe2SM}
\end{eqnarray}
(Note that I can extend the limits on the integral to $0$ and $1$ with impunity because of the aforementioned fact that  $z(e)$ is well behaved at those limits. )

The integral in this expression is elementary, and is
\begin{eqnarray}
\int_0^1 z(e')de'=\ln 2 -{870\over 2299}\ln\left({425\over304}\right)=.5663514....\,\,.
\label{int}
\end{eqnarray}

Using this result in (\ref{rpe2SM}) and solving for $e(r_p)$ gives
\begin{eqnarray}
e(r_p)=C_e\left({r_p\over r_{p0}}\right)^{19\over12}\,\,,
\label{rpe3SM}
\end{eqnarray}
where $C_e=\left({425\over304}\right)^{{145\over242}}/2^{{19\over12}}=.40790...$.
Using the relation (\ref{rmbSM1}) between the minimum first passage distance $r_{p0}$ and the 
impact parameter $b$, I can rewrite this as a relation between the eccentricity and the impact parameter:
\begin{eqnarray}
e(b)=C_e\left({r_Sr_p\over b^2}\cv^2\right)^{19\over12}\,\,.
\label{ebSM}
\end{eqnarray}
Solving for $b(e)$ gives
\begin{eqnarray}
b(e)=C_e^{6\over19}\sqrt{r_Sr_p}\cv e^{-{6\over19}}\,\,.
\label{beSM}
\end{eqnarray}
The probability distribution for the final eccentricity can now be obtained from that for the impact parameter $b$ via simple statistics, which imply:
\begin{eqnarray}
p(e;v_\infty)=p(b;v_\infty)\left|{db\over de}\right|\,\,.
\label{pe1SM}
\end{eqnarray}
Since the impact parameters of captured pairs should be uniformly distributed over a circle of radius $b_c$, I have
\begin{eqnarray}
p(b;v_\infty)={2b\over b_c^2(v_\infty)}\,\,.
\label{pbSM}
\end{eqnarray}
\newline
Using this and my expressions equation (\ref{beSM}) for $b(e)$ and (\ref{bc}) for $b_c(v_\infty)$ in (\ref{pe1}) gives equation (\ref{edist}). The lower limit $e_{min}$ on $e$ follows from recognizing that we must have $b<b_c$ if the pair are to be captured; therefore replacing $b$ with $b_c$ in equation (\ref{ebSM}) gives $e_{min}$, equation (\ref{emin}) of the main text.
I reiterate both of these equations here for completeness:

\begin{widetext}
\begin{eqnarray}
p(e;r_p)=\left\{
\begin{array}{ll}
C_p\left({r_p\over r_S(M)}\right)[f(1-f)]^{-{2\over 7}}\left({v_\infty\over c}\right)^{4\over 7} e^{-{31\over19}}\,,
&e_{min}<e\ll1\,,
\\ \\
0\,\,\,\,\,\,\,\,\,,
&e<e_{min}\,,
\end{array}\right.
\nonumber\\
\label{edistSM}
\end{eqnarray}
\end{widetext}

\begin{eqnarray}
e_{min}=C_{em}\left({r_p\over r_S(M)}\right)^{19\over12}[f(1-f)]^{-{19\over 42}}\left({v_\infty\over c}\right)^{19\over 21} \,\,,
\label{eminSM}
\end{eqnarray}
where 
\begin{eqnarray}
C_p={12 C_e^{12\over19}\over19C_b^2}={6\over19}\left({425\over304}\right)^{{870\over2299}}\left({96\over85\pi}\right)^{{2\over7}}=.2676222338...\nonumber
\label{C_pSM}
\end{eqnarray}
and
\begin{eqnarray}
C_{em}={C_e\over C_b^{19\over6}}=\left({425\over304}\right)^{{145\over242}}\left({96\over85\pi}\right)^{{19\over42}}2^{-{19\over12}}=
.25678305711...\nonumber\,.
\end{eqnarray} 
\newline

To predict the actual distribution of observations, for which the asymptotic approach velocity $v_\infty$ will of course be unknown, this must be averaged over the distribution of approach velocities.  That is,
\begin{eqnarray}
p(e)=\int_0^\infty dv_\infty \ p(v_\infty)p(e;v_\infty)\,\,.
\label{pevbarSM}
\end{eqnarray}

There is no contribution to $p(e)$ from velocities that are so large that $e_{min}(v_\infty)>e$. Equivalently, this means that the integral over $v_\infty$ in (\ref{pevbarSM}) has an upper limit $v_{max}$ determined by $e_{min}(v_{max})=e$. Solving this condition for $v_{max}$ implies that

\begin{eqnarray}
v_{max}=c\,C_v\left({r_S(M)\over r_p}\right)^{7\over4}\sqrt{f(1-f)}e^{21\over19} \,\,,
\label{vmaxSM}
\end{eqnarray}
where I've defined $C_v\equiv C_{em}^{-{21\over19}}=\left({425\over304}\right)^{-{3045\over4598}}\sqrt{85\pi\over96}2^{7\over4}=4.4934979...$

Taking the {\it speeds} $v_\infty$ to have a Maxwellian distribution:
\begin{eqnarray}
p(v_\infty)=\sqrt{{2\over\pi}}v_\infty^2 \exp\left(-v_\infty^2/ (2v_\sigma^2)\right)/ v_\sigma^3\,\,,
\label{pvSM}
\end{eqnarray}
which of course corresponds simply to taking the individual Cartesian components of $\bf{v}$
to have a Gaussian distribution, gives
\begin{widetext}
\begin{eqnarray}
p(e)&=&\int_0^{v_{max}(e)} dv_\infty \ p(v_\infty)p(e;v_\infty)\nonumber\\&=&\sqrt{{2\over\pi}}C_p\left({r_p\over r_S(M)}\right)[f(1-f)]^{-{2\over 7}}\left({1\over v_\sigma^3c^{4\over 7}}\right) e^{-{31\over19}}\int_0^{v_{max}(e)} dv \ v^{18/7}\exp\left(-v_\infty^2/ (2v_\sigma^2)\right)\,\,.
\label{pevbar2SM}
\end{eqnarray}
\end{widetext}
Changing variables of integration to $u\equiv{v^2\over2v_\sigma^2}$, and defining the characteristic eccentricity scale 
\begin{eqnarray}
e_c=C_{em}\left({r_p\over r_S(M)}\right)^{19\over12}[f(1-f)]^{-{19\over 42}}\left({v_\sigma\over c}\right)^{19\over 21} \,\,,
\label{echarSM}
\end{eqnarray}
I obtain
\begin{widetext}
\begin{eqnarray}
p(e)&=&\bar{C}\left(1\over e_c\right)\left(e_c\over e\right)^{31/19}\int_0^{{v_{max}^2(e)\over 2v_\sigma^2}} du \ u^{11/14}e^{-u}=\bar{C}\left(1\over e_c\right)\left(e_c\over e\right)^{31/19}\gamma\left({25\over14},{1\over 2}\left({e\over e_c}\right)^{42\over19}\right)\,\,,
\label{pevbarfSM}
\end{eqnarray}
\end{widetext}
where $\gamma(s, x)$ is the lower incomplete gamma function\cite{hurwitz}, and I've defined $\bar{C}\equiv2^{2\over7}\left({24\over19\sqrt{\pi}}\right)=.8687429...$.
This has the limiting forms:
\begin{eqnarray}
p(e)\approx
\left\{\begin{array}{ll}
{C_<\over e_c}\left({e\over e_c}\right)^{44\over19},& e\ll
e_c\,,\\ {C_>\over e_c}\left({e_c\over e}\right)^{31\over19},& e\gg
e_c\,,
\end{array}\right.
\label{pelimSM} 
\end{eqnarray}
where I've defined $C_<\equiv{168\over475\sqrt{2\pi}}=.141099585...$ and $C_>\equiv\bar{C}\Gamma\left({25\over14}\right)=.805906...$, where $\Gamma$ is the {\it complete} Gamma function. 

Of course, we do not really know what the distribution of asymptotic approach speeds $v_\infty$ is. Fortunately, the asymptotic form of the final eccentricity distribution for final eccentricities large compared to the typical eccentricities can be calculated with no knowledge of that distributiuon, as I'll now show.

For an arbitrary distribution $p(v_\infty)$ of asymptotic speeds, I have 
\begin{eqnarray}
p(e)&=&\int_0^{v_{max}(e)} dv_\infty \ p(v_\infty)p(e;v_\infty)\,\,.
\label{pevbargen1SM}
\end{eqnarray}
 
 For $e\gg e_{char}$, $v_m(e)\gg v_{char}$; therefore, I can take the upper limit on the integral in (\ref{pevbargen1SM}) to infinity. Doing so, and using my expression (\ref{edist}) for $p(e;v_\infty)$, I get
\begin{widetext}
\begin{eqnarray}
p(e\gg e_{char})&\approx&C_p\left({r_p\over r_S(M)}\right)[f(1-f)]^{-{2\over 7}} c^{-{4\over 7}} e^{-{31\over19}}\int_0^\infty dv_\infty \ p(v_\infty)v_\infty^{4/7}\,\,.
\label{pevbargen2SM}
\end{eqnarray}
\end{widetext}
The integral in this expression is simply $\left<v_\infty^{4\over7}\right>$. Hence, defining $v_c\equiv\left(\left<v_\infty^{4/7}\right>\right)^{7/4}$, I obtain 
\begin{widetext}
\begin{eqnarray}
p(e\gg e_{char})&\approx&C_p\left({r_p\over r_S(M)}\right)[f(1-f)]^{-{2\over 7}} \left({v_c\over c}\right)^{{4\over 7}} e^{-{31\over19}}\,\,,
\label{pevbargen2SM}
\end{eqnarray}
\end{widetext}
which is the general asymptotic result claimed in the main text.

\subsection{Distribution of Inspiral times}
 
 I note here that the distribution of inspiral times is extremely broad, as illustrated in the following table \ref{table:1}, which uses the masses of \gw, and takes  $v_\infty=30{\rm{km}\over\rm{sec}}$:
\begin{widetext}
\begin{center}

\section*{Gravitational radiation assisted capture inspiral times}

\end{center}

$b/b_c\,\,\,\,\,\,\,\,\,\,\,\,\,\,\,\,\,\,\,\,\,\,\,\,\,\,\,\,\,\,\,\,\,\,\,\,\,\,\,\,\,\,\,\,$  $P_<(b)\,\,\,\,\,\,\,\,\,\,\,\,\,\,\,\,\,\,\,\,\,\,\,\,\,\,\,\,\,\,\,\,\,\,\,\,\,\,\,\,$       $P_>(b)\,\,\,\,\,\,\,\,\,\,\,\,\,\,\,\,\,\,\,\,\,\,\,\,\,\,\,\,\,\,\,\,\,\,\,\,\,\,\,\,\,\,\,\,\,\,\,$  $T_{inspiral}$ (years)

\noindent\rule[0.25\baselineskip]{\textwidth}{1pt}

$.1\,\,\,\,\,\,\,\,\,\,\,\,\,\,\,\,\,\,\,\,\,\,\,\,\,\,\,\,\,\,\,\,\,\,\,\,\,\,\,\,\,\,\,\,\,\,\,\,\,\,$       $.01\,\,\,\,\,\,\,\,\,\,\,\,\,\,\,\,\,\,\,\,\,\,\,\,\,\,\,\,\,\,\,\,\,\,\,\,\,\,\,\,\,\,\,\,\,\,\,\,\,\,$$.99\,\,\,\,\,\,\,\,\,\,\,\,\,\,\,\,\,\,\,\,\,\,\,\,\,\,\,\,\,\,\,\,\,\,\,\,\,\,\,\,\,\,\,\,\,\,\,\,\,\,$        $5.37\times10^{-9} =(1/6) \rm{sec}$

$.2\,\,\,\,\,\,\,\,\,\,\,\,\,\,\,\,\,\,\,\,\,\,\,\,\,\,\,\,\,\,\,\,\,\,\,\,\,\,\,\,\,\,\,\,\,\,\,\,\,\,$         $.04\,\,\,\,\,\,\,\,\,\,\,\,\,\,\,\,\,\,\,\,\,\,\,\,\,\,\,\,\,\,\,\,\,\,\,\,\,\,\,\,\,\,\,\,\,\,\,\,\,\,$$.96\,\,\,\,\,\,\,\,\,\,\,\,\,\,\,\,\,\,\,\,\,\,\,\,\,\,\,\,\,\,\,\,\,\,\,\,\,\,\,\,\,\,\,\,\,\,\,\,\,\,$             $7.776\times10^{-6} = 4$ minutes

$.3\,\,\,\,\,\,\,\,\,\,\,\,\,\,\,\,\,\,\,\,\,\,\,\,\,\,\,\,\,\,\,\,\,\,\,\,\,\,\,\,\,\,\,\,\,\,\,\,\,\,$           $.09\,\,\,\,\,\,\,\,\,\,\,\,\,\,\,\,\,\,\,\,\,\,\,\,\,\,\,\,\,\,\,\,\,\,\,\,\,\,\,\,\,\,\,\,\,\,\,\,\,\,$$.91\,\,\,\,\,\,\,\,\,\,\,\,\,\,\,\,\,\,\,\,\,\,\,\,\,\,\,\,\,\,\,\,\,\,\,\,\,\,\,\,\,\,\,\,\,\,\,\,\,\,$           $5.493\times10^{-4} = 4.8$ hours

$.4\,\,\,\,\,\,\,\,\,\,\,\,\,\,\,\,\,\,\,\,\,\,\,\,\,\,\,\,\,\,\,\,\,\,\,\,\,\,\,\,\,\,\,\,\,\,\,\,\,\,$         $.16\,\,\,\,\,\,\,\,\,\,\,\,\,\,\,\,\,\,\,\,\,\,\,\,\,\,\,\,\,\,\,\,\,\,\,\,\,\,\,\,\,\,\,\,\,\,\,\,\,\,$$.84\,\,\,\,\,\,\,\,\,\,\,\,\,\,\,\,\,\,\,\,\,\,\,\,\,\,\,\,\,\,\,\,\,\,\,\,\,\,\,\,\,\,\,\,\,\,\,\,\,\,$            $1.12753\times10^{-2} = 4.12$ days

$.5\,\,\,\,\,\,\,\,\,\,\,\,\,\,\,\,\,\,\,\,\,\,\,\,\,\,\,\,\,\,\,\,\,\,\,\,\,\,\,\,\,\,\,\,\,\,\,\,\,\,$         $.25\,\,\,\,\,\,\,\,\,\,\,\,\,\,\,\,\,\,\,\,\,\,\,\,\,\,\,\,\,\,\,\,\,\,\,\,\,\,\,\,\,\,\,\,\,\,\,\,\,\,$$.75\,\,\,\,\,\,\,\,\,\,\,\,\,\,\,\,\,\,\,\,\,\,\,\,\,\,\,\,\,\,\,\,\,\,\,\,\,\,\,\,\,\,\,\,\,\,\,\,\,\,$            $ 0.118 = 43\,\, \rm{days} \approx 6$ weeks

$.6\,\,\,\,\,\,\,\,\,\,\,\,\,\,\,\,\,\,\,\,\,\,\,\,\,\,\,\,\,\,\,\,\,\,\,\,\,\,\,\,\,\,\,\,\,\,\,\,\,\,$           $.36\,\,\,\,\,\,\,\,\,\,\,\,\,\,\,\,\,\,\,\,\,\,\,\,\,\,\,\,\,\,\,\,\,\,\,\,\,\,\,\,\,\,\,\,\,\,\,\,\,\,$$.64\,\,\,\,\,\,\,\,\,\,\,\,\,\,\,\,\,\,\,\,\,\,\,\,\,\,\,\,\,\,\,\,\,\,\,\,\,\,\,\,\,\,\,\,\,\,\,\,\,\,$           0.813

$.7\,\,\,\,\,\,\,\,\,\,\,\,\,\,\,\,\,\,\,\,\,\,\,\,\,\,\,\,\,\,\,\,\,\,\,\,\,\,\,\,\,\,\,\,\,\,\,\,\,\,$          $.49\,\,\,\,\,\,\,\,\,\,\,\,\,\,\,\,\,\,\,\,\,\,\,\,\,\,\,\,\,\,\,\,\,\,\,\,\,\,\,\,\,\,\,\,\,\,\,\,\,\,$$.51\,\,\,\,\,\,\,\,\,\,\,\,\,\,\,\,\,\,\,\,\,\,\,\,\,\,\,\,\,\,\,\,\,\,\,\,\,\,\,\,\,\,\,\,\,\,\,\,\,\,$            4.3

$.8\,\,\,\,\,\,\,\,\,\,\,\,\,\,\,\,\,\,\,\,\,\,\,\,\,\,\,\,\,\,\,\,\,\,\,\,\,\,\,\,\,\,\,\,\,\,\,\,\,\,$          $.64\,\,\,\,\,\,\,\,\,\,\,\,\,\,\,\,\,\,\,\,\,\,\,\,\,\,\,\,\,\,\,\,\,\,\,\,\,\,\,\,\,\,\,\,\,\,\,\,\,\,$$.36\,\,\,\,\,\,\,\,\,\,\,\,\,\,\,\,\,\,\,\,\,\,\,\,\,\,\,\,\,\,\,\,\,\,\,\,\,\,\,\,\,\,\,\,\,\,\,\,\,\,$           19.7

$.8715\,\,\,\,\,\,\,\,\,\,\,\,\,\,\,\,\,\,\,\,\,\,\,\,\,\,\,\,\,\,\,\,\,\,\,\,\,\,\,\,\,$         $.76\,\,\,\,\,\,\,\,\,\,\,\,\,\,\,\,\,\,\,\,\,\,\,\,\,\,\,\,\,\,\,\,\,\,\,\,\,\,\,\,\,\,\,\,\,\,\,\,\,\,$$.24\,\,\,\,\,\,\,\,\,\,\,\,\,\,\,\,\,\,\,\,\,\,\,\,\,\,\,\,\,\,\,\,\,\,\,\,\,\,\,\,\,\,\,\,\,\,\,\,\,\,$          60

$.9\,\,\,\,\,\,\,\,\,\,\,\,\,\,\,\,\,\,\,\,\,\,\,\,\,\,\,\,\,\,\,\,\,\,\,\,\,\,\,\,\,\,\,\,\,\,\,\,\,\,$         $.81\,\,\,\,\,\,\,\,\,\,\,\,\,\,\,\,\,\,\,\,\,\,\,\,\,\,\,\,\,\,\,\,\,\,\,\,\,\,\,\,\,\,\,\,\,\,\,\,\,\,$$.19\,\,\,\,\,\,\,\,\,\,\,\,\,\,\,\,\,\,\,\,\,\,\,\,\,\,\,\,\,\,\,\,\,\,\,\,\,\,\,\,\,\,\,\,\,\,\,\,\,\,$            100

$.934\,\,\,\,\,\,\,\,\,\,\,\,\,\,\,\,\,\,\,\,\,\,\,\,\,\,\,\,\,\,\,\,\,\,\,\,\,\,\,\,\,\,\,\,$        $.87\,\,\,\,\,\,\,\,\,\,\,\,\,\,\,\,\,\,\,\,\,\,\,\,\,\,\,\,\,\,\,\,\,\,\,\,\,\,\,\,\,\,\,\,\,\,\,\,\,\,$$.13\,\,\,\,\,\,\,\,\,\,\,\,\,\,\,\,\,\,\,\,\,\,\,\,\,\,\,\,\,\,\,\,\,\,\,\,\,\,\,\,\,\,\,\,\,\,\,\,\,\,$           200

$.9771\,\,\,\,\,\,\,\,\,\,\,\,\,\,\,\,\,\,\,\,\,\,\,\,\,\,\,\,\,\,\,\,\,\,\,\,\,\,\,\,\,$       $.955\,\,\,\,\,\,\,\,\,\,\,\,\,\,\,\,\,\,\,\,\,\,\,\,\,\,\,\,\,\,\,\,\,\,\,\,\,\,\,\,\,\,\,\,\,\,\,$$.045\,\,\,\,\,\,\,\,\,\,\,\,\,\,\,\,\,\,\,\,\,\,\,\,\,\,\,\,\,\,\,\,\,\,\,\,\,\,\,\,\,\,\,\,\,\,\,$          1000

$.985617\,\,\,\,\,\,\,\,\,\,\,\,\,\,\,\,\,\,\,\,\,\,\,\,\,\,\,\,\,\,\,\,\,\,\,$       $.97\,\,\,\,\,\,\,\,\,\,\,\,\,\,\,\,\,\,\,\,\,\,\,\,\,\,\,\,\,\,\,\,\,\,\,\,\,\,\,\,\,\,\,\,\,\,\,\,\,\,$$.03\,\,\,\,\,\,\,\,\,\,\,\,\,\,\,\,\,\,\,\,\,\,\,\,\,\,\,\,\,\,\,\,\,\,\,\,\,\,\,\,\,\,\,\,\,\,\,\,\,\,$        2000

$.9921895\,\,\,\,\,\,\,\,\,\,\,\,\,\,\,\,\,\,\,\,\,\,\,\,\,\,\,\,\,\,\,\,$       $.984\,\,\,\,\,\,\,\,\,\,\,\,\,\,\,\,\,\,\,\,\,\,\,\,\,\,\,\,\,\,\,\,\,\,\,\,\,\,\,\,\,\,\,\,\,\,\,$$.016\,\,\,\,\,\,\,\,\,\,\,\,\,\,\,\,\,\,\,\,\,\,\,\,\,\,\,\,\,\,\,\,\,\,\,\,\,\,\,\,\,\,\,\,\,\,\,$      5000

$.9950703\,\,\,\,\,\,\,\,\,\,\,\,\,\,\,\,\,\,\,\,\,\,\,\,\,\,\,\,\,\,\,\,$      $.99\,\,\,\,\,\,\,\,\,\,\,\,\,\,\,\,\,\,\,\,\,\,\,\,\,\,\,\,\,\,\,\,\,\,\,\,\,\,\,\,\,\,\,\,\,\,\,\,\,\,$$.01\,\,\,\,\,\,\,\,\,\,\,\,\,\,\,\,\,\,\,\,\,\,\,\,\,\,\,\,\,\,\,\,\,\,\,\,\,\,\,\,\,\,\,\,\,\,\,\,\,\,$       10,000

$.9989314326\,\,\,\,\,\,\,\,\,\,\,\,\,\,\,\,\,\,\,\,\,\,\,$     $.998\,\,\,\,\,\,\,\,\,\,\,\,\,\,\,\,\,\,\,\,\,\,\,\,\,\,\,\,\,\,\,\,\,\,\,\,\,\,\,\,\,\,\,\,\,\,\,$$.002\,\,\,\,\,\,\,\,\,\,\,\,\,\,\,\,\,\,\,\,\,\,\,\,\,\,\,\,\,\,\,\,\,\,\,\,\,\,\,\,\,\,\,\,\,\,\,$    100,000

$.999769241\,\,\,\,\,\,\,\,\,\,\,\,\,\,\,\,\,\,\,\,\,\,\,\,\,\,$     $.9995\,\,\,\,\,\,\,\,\,\,\,\,\,\,\,\,\,\,\,\,\,\,\,\,\,\,\,\,\,\,\,\,\,\,\,\,\,\,\,\,\,\,\,\,$$.0005\,\,\,\,\,\,\,\,\,\,\,\,\,\,\,\,\,\,\,\,\,\,\,\,\,\,\,\,\,\,\,\,\,\,\,\,\,\,\,\,\,\,\,\,$      1,000,000 

$\,\,\,\,\,\,\,\,\,\,\,\,\,\,\,\,\,\,\,\,\,\,\,\,\,\,\,\,\,\,\,\,\,\,\,\,\,\,\,\,\,\,$     $\,\,\,\,\,\,\,\,\,\,\,\,\,\,\,\,\,\,\,\,\,\,\,\,\,\,\,\,\,\,\,\,\,\,\,\,\,\,\,\,\,\,\,\,\,\,\,\,\,\,\,\,\,\,\,\,\,\,\,\,$$\,\,\,\,\,\,\,\,\,\,\,\,\,\,\,\,\,\,\,\,\,\,\,\,\,\,\,\,\,\,\,\,\,\,\,\,\,\,\,\,\,\,\,\,\,\,\,\,\,\,\,\,\,\,\,\,\,\,\,\,$     \,\,\,\,\,\,\,\,\,\,\,\,\,\,\,\,

\end{widetext}

 \noindent Here $P_{_{_<}}(b)=\left(b/b_c\right)^2$ and $P_{_{_>}}(b)=1-P_{_{_<}}(b)$ denote 
 \newline
 the probabilities of impact parameters less than $b$, and greater than $b$, respectively.
 
Note that inspirals in the bottom four percentile take less than four minutes, while those in the top four percentile take more than $1000$ years! Note, however, that none of these times is cosmologically significant.\newline
 
It is not possible to get a closed form analytic expression for the distribution of inspiral times, due to the impossibility of analytically inverting the Hurwitz zeta function in my expression (3) of the main text for the inspiral time. It is, however, possible to obtain this distribution in the limits of inspiral times $\tau\ll\tau_{_{med}}$ and $\tau\gg\tau_{_{med}}$.

In the former limit, which corresponds to $b\ll b_c$, the argument $x$ of the Hurwitz zeta function goes to $1$, and the Hurwitz zeta function goes to a constant, namely, the Riemann zeta function $\zeta({3\over2})$. My expression (3) of the main text for the inspiral time then reduces to 
\begin{eqnarray}
\tau&\approx&\pi\left({c^2 r_S(M)\over v_\infty^3}\right)\left({b\over b_c}\right)^{21/2}\zeta\left({3\over 2}\right) \,.
\label{inspiralsmalltau}
\end{eqnarray}
The cumulative probability $P_{_{_<}}(\tau)$ that the inspiral time is less than some specified $\tau$ is just 
$\left({b\over b_c}\right)^2$; solving (\ref{inspiralsmalltau}) for that quantity gives
\begin{eqnarray}
P_{_{_<}}(\tau\ll\tau_{_{med}})&=&\left({b\over b_c}\right)^2=\left({\tau v_\infty^3\over\pi c^2 r_S(M) \zeta\left({3\over 2}\right)}\right)^{4\over21} \,.
\label{inspiralsmalltau}
\end{eqnarray}
This can conveniently be written in terms of the median inspiral time $\tau_{_{med}}$, which is just $\tau$ evaluated from (\ref{inspiral}) at the median impact parameter $b_{med}=b_c/\sqrt{2}$; this gives
\begin{eqnarray}
\tau_{_{med}}&=&C_{med}\left({\pi c^2 r_S(M)\over v_\infty^3}\right)
\label{taumed}
\end{eqnarray}
where I've defined
\begin{eqnarray}
C_{med}\equiv
2^{-{21/4}}\zeta\left({3\over 2}, 1-2^{-{7\over2}}\right)\approx.073802 \,.
\label{Cmed}
\end{eqnarray}
Using this in (\ref{inspiralsmalltau}) gives
\begin{eqnarray}
P_{_{_<}}(\tau)&=&C_{_{\tau<}}\left({\tau\over\tau_{_{med}}}\right)^{4\over21} \,,
\label{P<}
\end{eqnarray}
where I've defined 
\begin{eqnarray}
C_{_{\tau<}}\equiv \left({C_{med}\over\zeta\left({3\over 2}\right)}\right)^{4\over21}={1\over2}\left({\zeta\left({3\over 2}, 1-2^{-{7\over2}}\right)\over\zeta\left({3\over 2}\right)}\right)^{4\over21}= .506942...\nonumber
\end{eqnarray}
For $\tau\gg\tau_{_{med}}$, 
which corresponds to $b\rightarrow b_c$, the argument $x=1-\left({b\over b_c}\right)^7$ of the Hurwitz zeta function in (\ref{inspiral}) is well approximated by $x\approx 7\epsilon$,  where $\epsilon\equiv 1-{b\over b_c}$, which  goes to $0$ as $b\rightarrow b_c$. In this limit,  the Hurwitz zeta function approaches $x^{-{3\over2}}$, while $\left({b\over b_c}\right)^{21\over2}\rightarrow1$. Putting these facts together in (\ref{inspiral}), I get a good approximation to $\tau$ for $\tau\gg\tau_{_{med}}$:
\begin{eqnarray}
\tau&\approx&{\pi\over7^{3\over2}}\left({c^2 r_S(M)\over v_\infty^3}\right)\epsilon^{-{3/2}} \,.
\label{inspiralbigtau}
\end{eqnarray}
\newline
In this limit, the cumulative probability $P_{_{_>}}(\tau)$ that the inspiral time is greater than some specified $\tau$ is just 
$1-\left({b\over b_c}\right)^2\approx2\epsilon$; solving (\ref{inspiralbigtau}) for $\epsilon$ then gives gives
\begin{eqnarray}
P_{_{_>}}(\tau)&=&{2\left(\pi c^2 r_S(M)\right)^{2\over3}\over 7v_\infty^2\tau^{2\over3}}=C_{_{\tau>}}\left({\tau_{_{med}}\over \tau}\right)^{2\over3} \,,
\label{P>}
\end{eqnarray}
where I've used 
my earlier result (\ref{taumed}) for the median time $\tau_{_{med}}$, and I've defined 
\begin{eqnarray}
C_{_{\tau>}}\equiv{2^{9\over2}\over7\left(\zeta\left({3\over 2}, 1-2^{-{7\over2}}\right)\right)^{2\over3}}=1.623877...
\label{CT>}
\end{eqnarray}

The small $\tau$ limit equation (\ref{P<}) is accurate to $2\%$ up to $\tau=.9\tau_{med}$, while the large $\tau$ limit (\ref{P>}) is accurate to $2\%$ down to $\tau=200\tau_{med}$; these two ranges contain more than half of all captures.
\newline

\subsection{Connection to earlier work of Walker and Will}







The first recognition of the possibility of GRAC was by Walker and Will\cite{Will}. In this subsection, I recover their result for the maximum {\it incoming} eccentricity of a pair that are captured; that is, in my terminology, the incoming eccentricity of two bodies approaching with impact parameter $b=b_c$.

Denoting the periastron distance of the two objects  on their first passage as $r_p$, the maximum value $r^c_p$ that $r_p$ can take on
(which occurs when the two objects approach each other with exactly the maximum impact parameter $b_c$ for capture) is given by
\begin{eqnarray}
r_p^c=\left({85\pi b_c^2f(1-f)\over 96r^2_s(M)}\right)^{2\over9} r_s(M)\,.
\label{rminSM}
\end{eqnarray}

This result can be used to determine the eccentricity of the orbit before capture, which can be compared with the result of \cite{Will}. To do so, I begin with the simple geometrical observation that
\begin{eqnarray}
b_c&=& \lim_{\theta\to\theta_c}{p_c\sin(\theta_c-\theta)\over 1+e\cos(\theta)}\nonumber\\
&=&2r^c_p\lim_{\theta\to\theta_c}{\theta_c-\theta\over\sin(\theta_c)(\theta_c-\theta)}={2r^c_p\over \sin(\theta_c)}\,,
\label{eccent1}
\end{eqnarray}
where $\theta_c=\arccos\left(-{1\over e}\right)$ is the angle at which $r\rightarrow\infty$, with $e$ the eccentricity of the orbit and $p$ its semilatus rectum. In writing (\ref{eccent1}), I have used the fact that, for a nearly parabolic orbit, $p\approx2r_p$. Another property of  a nearly parabolic orbit is that $e=1+\epsilon$ with $\epsilon\ll1$, which is the case for any two bodies that are captured by GRAC when $v_\infty\ll c$. In this case, it is straightforward to show that $\sin(\theta_c)\approx\sqrt{2\epsilon}$. Using this in (\ref{eccent1}), I find
\begin{eqnarray}
b_c=r^c_p\sqrt{{2\over\epsilon}}\,.
\label{eccent2}
\end{eqnarray}
Now solving (\ref{rminSM}) to write $b_c$ in terms of $r^c_p$ on the right hand side of this expression, and solving the resultant equation for $\epsilon$ gives
\begin{eqnarray}
\epsilon=(170\pi/3)\left({\mu\over M}\right)\left({r_S(M)\over 2p_c}\right)^{5\over 2}\,,
\label{eccent3}
\end{eqnarray}
where $\mu \equiv{m_1m_2\over M}=f(1-f)M$ 
is the reduced mass. 
Equation (\ref{eccent3}) is the final (unnumbered) equation of reference\cite{Will} (which uses "natural" units in which $c=1=G$; in those units, $r_S(M)=2M$). In writing 
(\ref{eccent3}), I have again used the fact that for a nearly parabolic orbit $p=2r_p$.


\begin{thebibliography}{99}

\bibitem{GWD1} B. P. Abbott {\it et. al.}, Phys. Rev. Lett. {\bf 116}, 061102 (2016).

\bibitem{GWD2} B. P. Abbott {\it et. al.}, Phys. Rev. Lett. {\bf 116}, 241102 (2016).

\bibitem{mtw}
C. W. Misner, K. S. Thorne, and J. A. Wheeler,  {\it Gravitation},
(Freeman 1973).

\bibitem{3 body} K. Belczynski, S. Repetto, D. Holz, R. O' Shaughnessy, T.
Bulik, E. Berti, C. Fryer, M. Dominik, Astophys. J. {\bf 819} 108 (2016);
S. Sigurdsson and L. Hernquist, Nature (London) {\bf 364}, 423
(1993); S. F. Portegies Zwart and S. L.W. McMillan, Astrophys. J.
Lett. {\bf 528}, L17 (2000).
[107] C. L. Rodriguez, M. Morscher, B. Pattabiraman, S.
Chatterjee, C.-J. Haster, and F. A. Rasio, Phys. Rev. Lett.
{\bf 115}, 051101 (2015).


\bibitem{tidal} Unless one considers tidal forces. However, these clearly cannot play a role in binding two black holes, since they are point masses.

\bibitem{Will} The possibility of such gravitational radiation assisted capture was first recognized by M. Walker and C. M. Will, Phys. Rev. D {\bf 19}, 3483 (1979).





\bibitem{hurwitz} M. Abramowitz and I. A. Stegun, {\it Handbook of Mathematical Functions}, (Dover Publications, New York 1964).

\bibitem{SM} See Supplemental Material at [URL will be inserted by publisher] for a detailed derivation of this result.

\bibitem{frey} R. Frey, private communication.

\bibitem{GBH} M. J. Rees and D. Lynden-Bell, Mon. Not. R. Astron. Soc. {\bf 152}, 461 (1971).

	
\bibitem{Milky GBH} S. Gillessen, F. Eisenhauer, S. Trippe, T. Alexander, R. Genzel, F. Martins, T. Ott,  Astrophys. J.,  {\bf 692}, 1075 (2009).

\bibitem{GBH2} S. G. Djorgovski,  M. Volonteri, V. Springel, V. Bromm, and G. Meylan, 
in {\it The Eleventh Marcel Grossmann Meeting on Recent
Developments in Theoretical and Experimental General Relativity,
Gravitation and Relativistic Field Theories}, ed. H. Kleinert,
R. T. Jantzen, and R. Ruffini (Singapore: World Scientific), pg. 340 (2008). 




\bibitem{Peters} P. C. Peters, Phys. Rev.  {\bf 136}, 1224 (1964).

\bibitem{Kepler} J. Kepler, {\it Harmonices Mundi}  (Linz, (Austria): Johann Planck, 1619).





\bibitem{binormal}  W. R. Hamilton, Proceedings of the Royal Irish Academy {\bf 3},  344 (1847). 

\bibitem{Newton} {\it Philosophi¾ Naturalis Principia Mathematica}, (Roy. Soc. London, 1687).

\end{thebibliography}
\end{document}